\documentclass[conference, a4paper]{IEEEtran}
\IEEEoverridecommandlockouts
\usepackage{cite}
\usepackage{amsmath,amssymb,amsfonts}
\usepackage{algorithmic}
\usepackage{graphicx}
\usepackage{textcomp}
\usepackage{xcolor}
\usepackage{dsfont}
\usepackage{amsmath}
\usepackage{enumitem}
\usepackage{fancyhdr}
\usepackage{booktabs}
\usepackage{caption}
\usepackage{colortbl}
\usepackage{xcolor}
\usepackage{multirow}
\def\BibTeX{{\rm B\kern-.05em{\sc i\kern-.025em b}\kern-.08em
    T\kern-.1667em\lower.7ex\hbox{E}\kern-.125emX}}
\begin{document}

\title{Guidelines for 5G End to End Architecture and Security Issues \\
}

\author{\begin{tabular}[t]{c@{\extracolsep{1em}}c}
		\centering Ta-Hao Ting ${}^{1}$, & Tsung-Nan Lin ${}^{2}$ \\
			\centering Shan-Hsiang Shen  ${}^{3}$, & Yu-Wei Chang ${}^{4}$ \\
	\end{tabular}
	{} \\
	\\
	\small ${}^1$ \emph{Graduate Institute of Communication Engineering,}   \\
	\small \emph{National Taiwan University,}   \\
	\small \emph{Taipei, Taiwan,} \\
	\small \emph{d07942008@ntu.edu.tw} \\\\
	
		\small ${}^2$ \emph{Graduate Institute of Communication Engineering,}   \\
	\small \emph{National Taiwan University,}   \\
	\small \emph{Taipei, Taiwan,} \\
	\small \emph{tsungnan@ntu.edu.tw} \\\\
	
	\begin{tabular}[t]{c@{\extracolsep{1em}}c}
		\small ${}^3$ \emph{Department of Computer Science and Information Engineering,}   \\
		\small \emph{National Taiwan University of Science and Technology,}   \\
		\small \emph{Taipei, Taiwan,} \\
		\small \emph{sshen@csie.ntust.edu.tw } \\
		{} \\
	\end{tabular}
		\begin{tabular}[t]{c@{\extracolsep{1em}}c}
		\small ${}^4$ \emph{Graduate Institute of Communication Engineering,}   \\
		\small \emph{National Taiwan University,}   \\
		\small \emph{Taipei, Taiwan,} \\
		\small \emph{b01505025@g.ntu.edu.tw} \\
		{} \\
	\end{tabular}
	\\
	\small \emph{} }

\maketitle

\maketitle
\begin{abstract}
	
Hackers target their attacks on the most vulnerable parts of a system. A system is therefore only as strong as its weakest part, similar to the Cannikin Law, which states that the capacity of a barrel of water depends on the height of the shortest rather than the longest piece of wood.

To ensure the security of 5G networks, we first need to understand the overall 5G architecture and examine the potential threats instead of merely setting up a firewall. However, 5G networks will have tremendous coverage. The development of 5G techniques require extensive resources, leading to intense competition between countries attempting to develop 5G networks. Many outstanding papers discuss the techniques for developing specific aspects of the 5G architecture, but to the best of our knowledge, few provide an overview of a complete 5G network.

This presents us with a difficult situation because we need to consider the overall architecture to ensure the security of 5G networks. To address that problem, in this paper we provide essential guidelines for understanding the architecture of 5G. 

We introduce 5G scenarios, outline the network architecture, and highlight the potential security issues for the various components of a 5G network.
This paper is intended to facilitate a preliminary understanding of the 5G architecture and the possible security concerns of the various components.

The end to end (E2E) 5G network architecture is composed of a next-generation radio access network (NG-RAN), multi-access edge computing (MEC), virtual evolved packet core (vEPC), a data network (DN) and a cloud service. Network slicing (NS), network function virtualization (NFV), NFV Management and Orchestration (MANO) and software-defined networking (SDN) are also important techniques for achieving 5G network architectures.
\end{abstract}

\section{Introduction}
5G aims to use the same network infrastructure to provide customized services for each vertical industry. 3GPP summarizes these vertical industries and classifies 5G scenarios into three categories:
\begin{itemize}
	\item Enhanced mobile broadband (eMBB) focuses on high data rate and high data traffic. 
	\item Ultra-reliable and low latency communications (URLLC) focuses on low latency, low error rate, and ultra-reliability.
	\item Massive Machine Type Communications (mMTC) focuses on massive numbers of connections and saving energy.
\end{itemize}


We introduce the technologies required to accomplish these demands under the same infrastructure.
\begin{itemize}
\item Network Slicing (NS): NS is a macroscopic concept that virtually separates the infrastructure into many parallel slices. The slices are isolated from each other to prevent congestion and to prevent faults in one slice from affecting the others. The concept helps 5G to provide a guaranteed service because current network services can only provide a "best effort," with no guarantee.
\item Next Generation Radio Access Network (NG-RAN)
5G radio can utilize the spectrum from 400 MHz to 100 GHz, and this is divided into remote radio heads (RRHs) and virtualized baseband units (BBUs). The base station is separated into two parts because 5G uses a higher spectrum than 4G; a higher spectrum means higher signal attenuation, so separating the base station into two parts allows better control.
\item Multi-Access Edge Computing (MEC):
To reduce the latency, some network functions are moved from the vEPC to the MEC. MEC can be considered a light vEPC.
\item virtual Evolved Packet Core (vEPC):
The vEPC virtualizes the LTE EPC, divides the original network functions into several parts and adds new functions.
\item Software-Defined Networking (SDN):
SDN is used to assist in the isolation and flexibility of NS.
\item Network Function Virtualization (NFV):
 A virtualized network function such as the vEPC's network function or a virtual firewall.
\item NFV Management and Orchestration (MANO):
A virtualization platform that manages many NFVs.
\end{itemize}

\begin{figure}[!h]
	\centering \centerline{
		\includegraphics[width=0.45\textwidth]{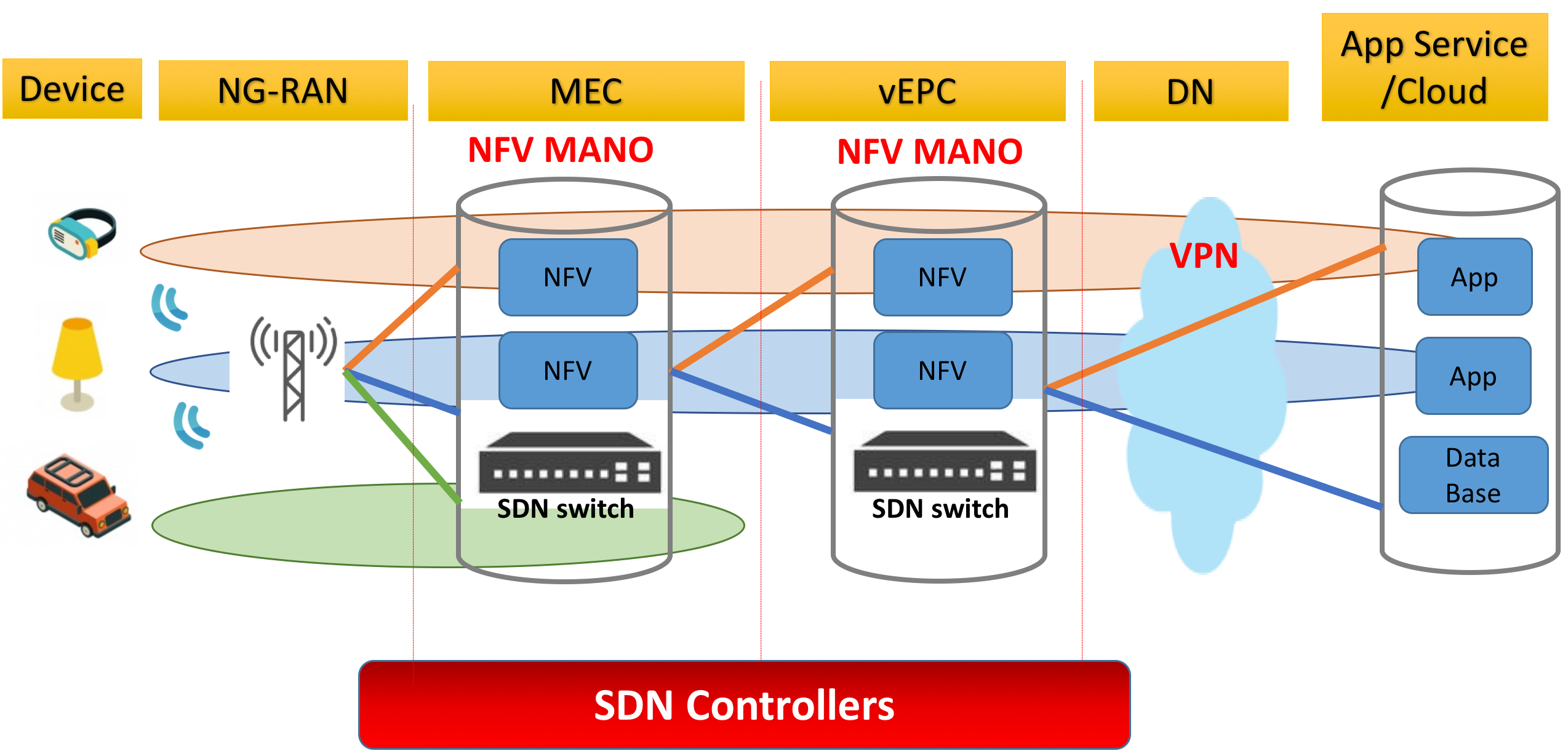}}
	\caption {5G E2E Network Architecture: the components of 5G are NG-RAN, MEC and vEPC.}
	\label{fig:E2E}
\end{figure}

The 5G E2E network architecture is illustrated in Fig. \ref{fig:E2E}, which provides an overview of 5G E2E networks. Fig.  \ref{fig:E2E}, Fig . \ref{fig:NS}, and Fig. \ref{fig:5GS} all show the same picture but emphasize different aspects.
The transition from 3G to 4G focused on the base station because wireless communication from the base station was the bottleneck. When progressing from 4G to 5G, however, the E2E architecture is more important because the base station is not the main bottleneck in the 5G network.
The 5G technologies mentioned above are described in detail in the next section.
\section{5G Architectural Framework}
\subsection{Network Slicing Architecture}
Network slicing provides a variety of independent service level agreements (SLA) for different needs. The services do not interfere with each other.

\begin{figure}[!h]
	\centering \centerline{
		\includegraphics[width=0.4\textwidth]{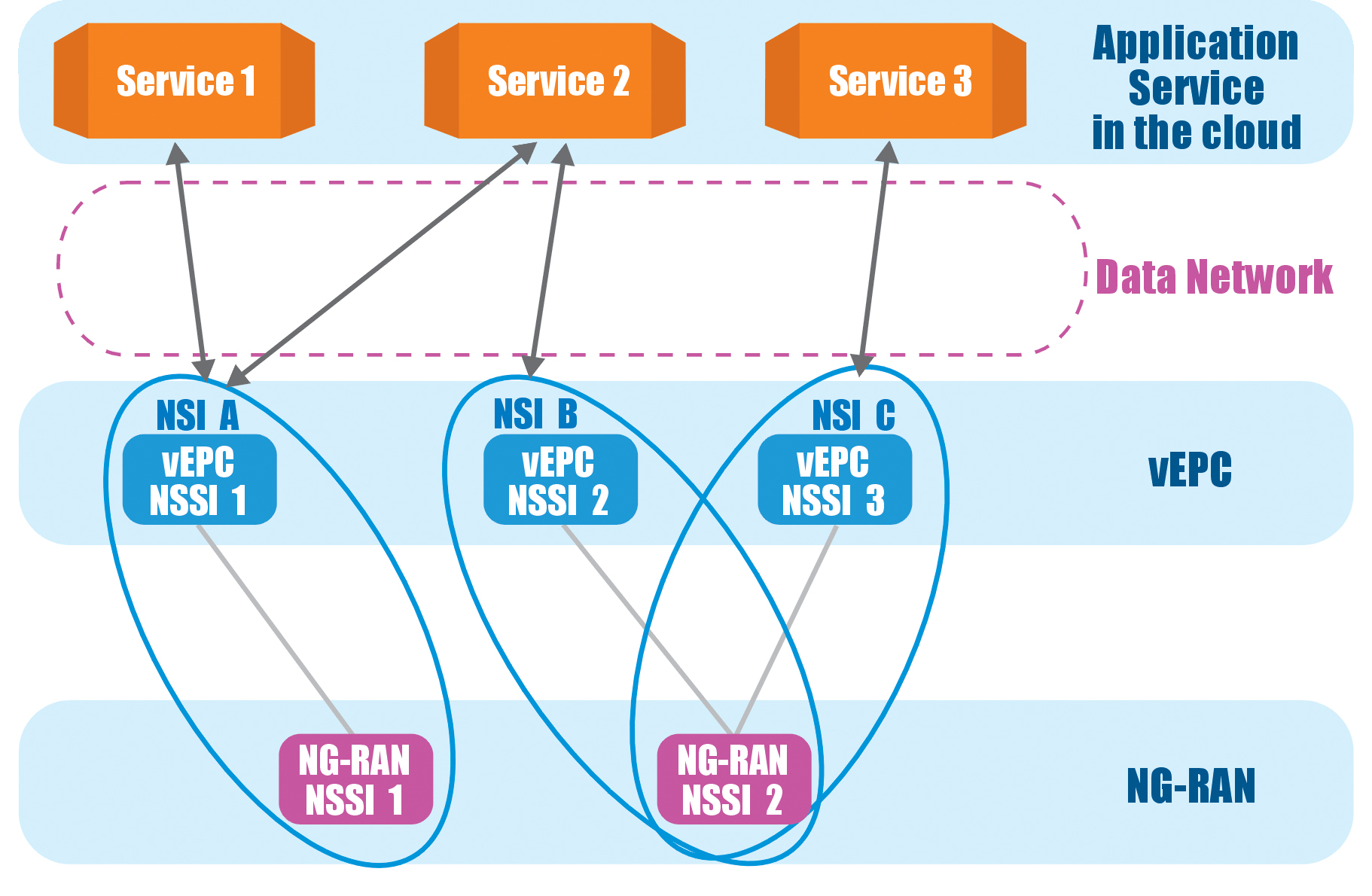}}
	\caption {5G Network Slicing Architecture, showing that an NSSI may provide several services at the same time but these services are isolated from each other.}
	\label{fig:NS}
\end{figure}

A network slice can be divided into two subsets: one for the RAN network slice subnet instance (NSSI) and the other for the CN NSSI. A slice can serve several different services at the same time, and the network slicing and services are independent. The network slicing may not disappear at the end of the service, so it is very flexible.
A schematic diagram of NS is shown in Fig. \ref{fig:NS}. The NS properties have previously been addressed in the 3GPP specification\cite{b3} \cite{b4} \cite{b5}.
The 3GPP specification is introduced below.

\subsubsection{The NS lifecycle consists of four steps}
\begin{itemize}		
	\item Preparation phase.
	\item Instantiation, Configuration and Activation phase.
	\item Run-time phase.
	\item Decommissioning phase.
\end{itemize}

\subsubsection{Types of Network Slicing}
\begin{itemize}	
	\item Shared constituent network slice instance (NSI): Many NSIs share one NSSI.
	\item Shared constituent NSSI: Many NSSIs share one network function(NF).
	\item Non-shared NSSI: An NSSI is used for only one NSI.
\end{itemize}

\subsubsection{Network Slicing Behavior}	
\begin{itemize}	
	\item An NSI may be created for one or more communication service. 
	\item The NSI and the communication service are independent; the NSI might not stop when a communication service is over because another communication service might start.
\end{itemize}

\subsection{Next-Generation Radio Access Network (NG-RAN)}
LTE base stations (BSs) are independent of each other. The disadvantage of this is that the wireless spectrum resources are seriously wasted. The mobile device will interfere with the surrounding BS interval. 

To overcome these problems in the LTE, the cloud radio access network has been defined and chosen as the next generation RAN (NG-RAN)\cite{b6} \cite{b7}. The NG-RAN system consists of RRHs and a pool of BBUs, and two connected front-haul networks. The RRHs collect wireless signals from mobile devices and is transmitted to the BBU pool with front-haul assistance. The virtualized and centrally controlled BBU pool can manage multiple base stations simultaneously and can dynamically allocate spectrum, time and spatial to meet dynamic traffic demands. The location of the NG-RAN in the 5G network is shown in Fig. \ref{fig:5GS}.

\subsection{Virtual Evolved Packet Core (vEPC) Network Functions }
\begin{figure}[!h]
	\centering \centerline{
		\includegraphics[width=0.5\textwidth]{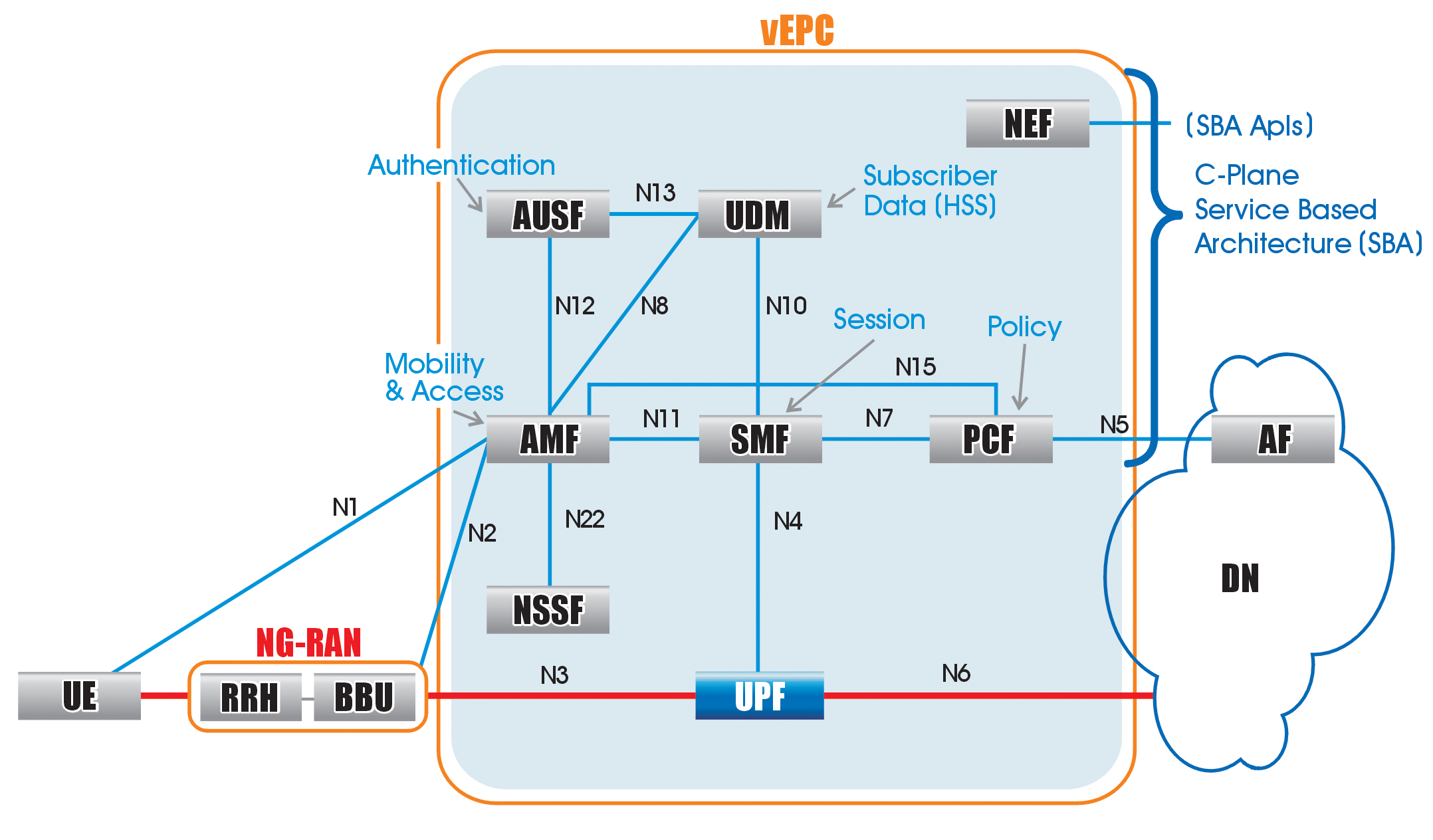}}
	\caption {5G vEPC architecture, focusing on the VNFs of a vEPC.}
	\label{fig:5GS}
\end{figure}

LTE EPC functions are also virtualized into multiple virtual network functions (VNFs). Virtualization offers the ability to quickly deploy service environments and reduce construction costs.
Several vEPC components are discussed below and shown in Fig.\ref{fig:5GS}. These components are compared with the functions of LTE EPCs.

\subsubsection{Access and Mobility Management Function (AMF)}
The AMF has some of the functions of the 4G EPC MME; it manages registration, connection, mobility, context, access authentication, and authorization.
\subsubsection{Session Management function (SMF)}
The SMF has some of the functions of the 4G EPC MME/PGW function; it manages the session, IP address allocation (static or DHCP), traffic steering, and policy enforcement.
\subsubsection{User Plane Function (UPF)}
The UPF has some of functions of the 4G EPC SGW/PGW; it manages packet routing and packet forwarding. The UPF also plays an important role in connecting to the DN and RAN.
\subsubsection{Policy Control Function (PCF)}
The PCF has some of the functions of the 4G EPC PCRF; it provides the policy rules to control the plane and unifies the policy framework.
\subsubsection{Authentication Server Function (AUSF)}
The AUSF has some of the function of the 4G HSS; it acts as the authentication server.
\subsubsection{Unified Data Management (UDM)}
UDM has some of the 4G EPC HSS function; it manages the user identification, subscriber data and authentication credentials.
\subsubsection{Network Exposure function (NEF)}
The NEF is new and does not appear in the 4G EPC. It provides an exposure interface to exchange information between internal and external networks.
\subsubsection{Application Function (AF)}
The AF has the same ability as the 4G AF function; it accesses the NEF and interacts with the PCF.
\subsubsection{Network Slice Selection Function (NSSF)}
NSSF is a new function that does not appear in the 4G EPC. It selects the proper network slice instances and AMF for the user.

\subsection{Software Defined Network (SDN)}
The inclusion of NFV MANO systems in 5G networks will inevitably increase the burden on the infrastructure. It is difficult to configure the device settings for such dynamic systems. The target of SDN is to use programmable software-driven devices to control the behavior of the infrastructure \cite{b8}. With SDN, 
user services can be delivered faster and the efficiency of network resources increases.
SDN is based on three principles:
\begin{itemize}
	\item decoupling of control from traffic forwarding and processing,
	\item logically centralized control, and
	\item programmability of network services.
\end{itemize}

\section{5G Management Platform}
Both NG-RAN and vEPC have been virtualized into many NFVs. The NFV system is larger and more flexible than the previous business model for telecom operators. The implementation of an NFV system involves combining the VNF into a network function chain, which evaluates how to forward packet flows from this VNF to another.
An efficient management solution is needed for the complicated architecture of 5G networks. The aim of NFV MANO systems is to process network resources, handle running services, decrease deployment costs and reduce the complexity of implementing new services.\\
\subsection{NFV MANO (Management and Orchestration)}
\begin{figure}[!h]
	\centering \centerline{
		\includegraphics[width=0.5\textwidth]{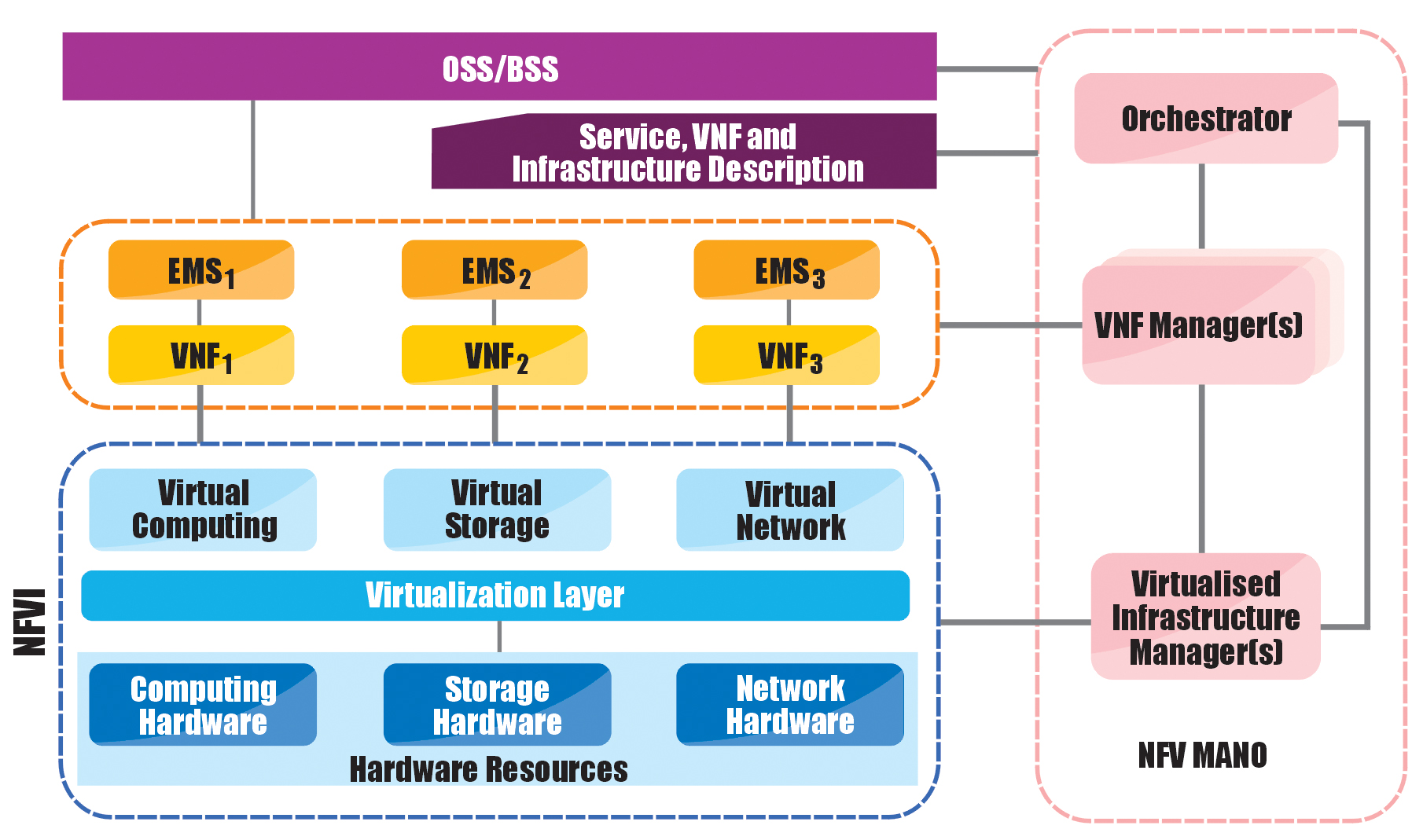}}
	\caption {ETSI NFV Architectural Framework}
	\label{fig:mano}
\end{figure}

The European Telecommunications Standards Institute (ETSI) published the NFV system architecture framework \cite{b9}, known as NFV management and orchestration (MANO) and shown in Fig.\ref{fig:mano}. 

Many open source organizations have developed their own NFV MANO frameworks, although they are basically based on the ETSI framework.
The modules and functions are as follows.

\subsubsection{NFV Infrastructure(NFVI)}Infrastructure for NFV technology that provides virtualized computing, storage, and networking.

\subsubsection{Virtual Network Function (VNF)}
The virtual network functions running on the NFVI, such as the firewall, NAT, and CDN.

\subsubsection{Operations Support System/Business Support System (OSS/BSS)}
Responsible for telecommunications services such as billing and user data maintenance.
\subsubsection{Virtual Infrastructure Manager (VIM)}
Responsible for managing NFVI and allocating virtual resources. OpenStack is often used to implement the VIM.
\subsubsection{VNF Manager(VNFM)}
Responsible for managing the VNF, its tasks include starting and stopping the VNF, status monitoring, and configuration settings.
\subsubsection{Network Function Virtualization Orchestrator (NFVO)}
Responsible for coordinating the VNFM and VIM according to the requirements of the OSS/BSS to orchestrate a specific service such as a firewall or to detect intrusions.
We introduced the architecture of the ETSI NFV system framework above and the combination of the VIM, VNFM, NFVO is referred to as NFV MANO. 

\subsection{NFV MANO Implementations}
Many open-source and commercial projects reference the ETSI framework model. We provide a brief introduction to these projects.

\subsubsection{OSM} Open-source MANO (OSM) is an ETSI project\cite{b17}.
OSM consists of three main software components, which can be mapped onto the ETSI MANO framework: 1. Virtual Infrastructure Management (VIM), 2. VNF Management, and 3. the NFV Orchestration (NFVO) layer.
\subsubsection{SONATA}
The SONATA\cite{b18} service platform is implemented in a modular micro-service, which is very flexible and helps the operator to modify a customized function. \cite{b19} integrates the network controller and the virtual machine, which are already implemented in SONATA. 

\subsubsection{Open NFV (OPNFV)}
Launched by the Linux Foundation in September 2014 \cite{b10}, OPNFV focuses on NFVI and includes an SDN controller and switch. OPNFV is a platform for implementing NFV and can feedback the necessary information to the ONAP platform. The four goals of OPNFV are as follows.
\begin{itemize}
	\item After testing, use continuous integration/continuous delivery (CI/CD) to develop an open source platform that can build NFV system functions.
	\item Invite telecommunication service operators to join the program and verify that OPNFV meets user expectations.
	\item Join other open source projects that will use OPNFV to ensure consistency, performance, and interoperability.
	\item Build an ecosystem of NFV solutions based on open source. 
\end{itemize}
\subsubsection{ONAP}
\begin{figure}[!h]
	\centering \centerline{
		\includegraphics[width=0.5\textwidth]{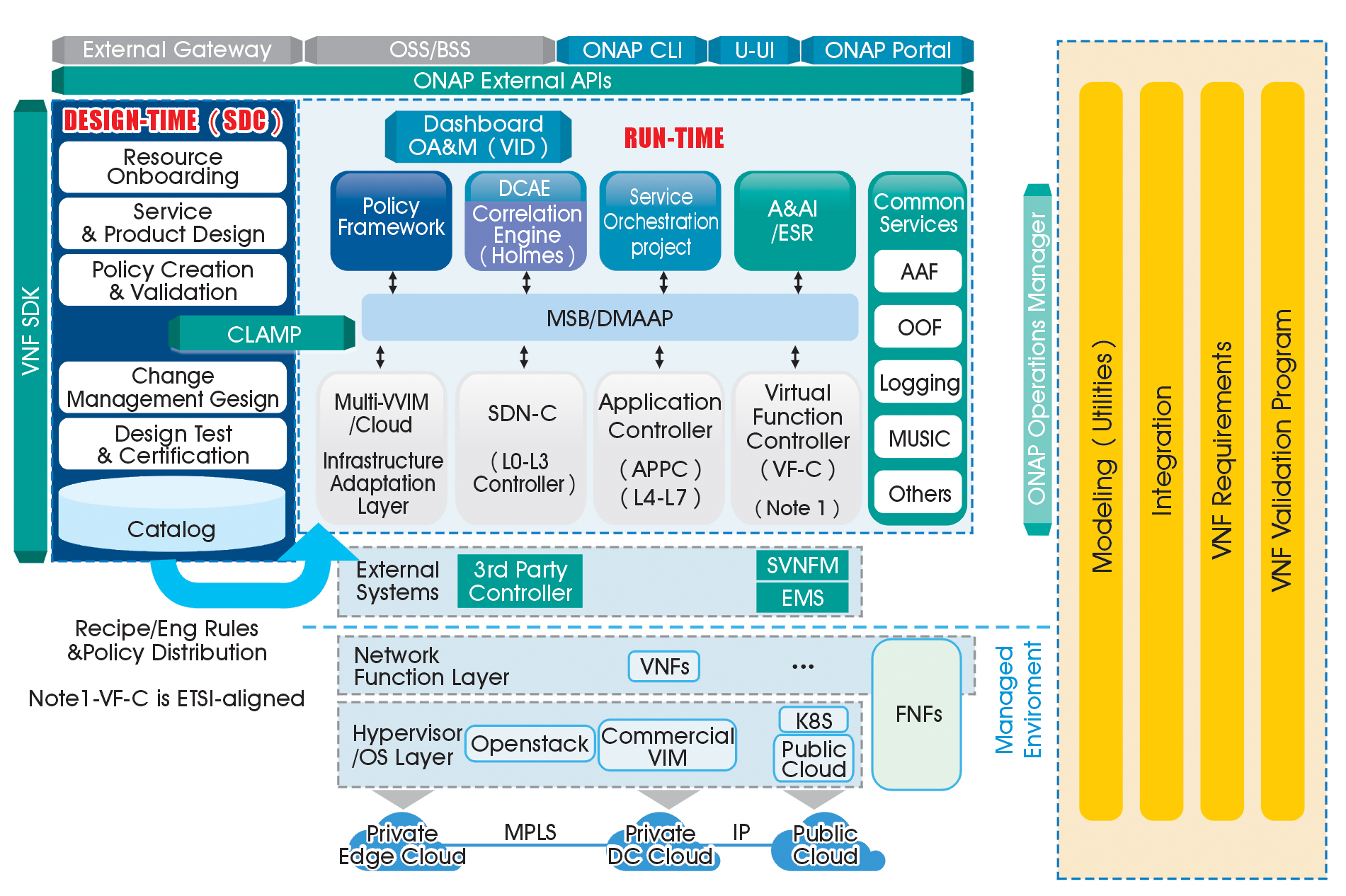}}
	\caption {ONAP Platform Components}
	\label{fig:ONAP}
\end{figure}

Formed in March 2017 \cite{b11}, the ONAP platform provides a real-time, policy-driven orchestration platform. It allows developers to automate new services and manage their lifecycles. The ONAP platform components are shown in Fig.\ref{fig:ONAP}. There are three major requirements.
\begin{itemize}
\item A robust design framework models the resources and relationships, guides the service behavior using policy rules, and flexibly manages the applications and closed-loop events.\\
\item An orchestration and control framework (Service Orchestrator and Controllers) instantiates the service automatically and manages service demands. \\
\item An analytic framework, based on the specified design, analyzes, monitors, and elastically adjusts the service behavior during the service lifecycle.
\end{itemize}

Controllers, Data Collection, Analytics and Events (DCAE) is a component that can help to extend the ONAP to provide functionalities related to security.  The DACE component is introduced below.
\begin{itemize}
	\item DCAE uses many analytic micro-services to collect usage, performance, and configuration data about the managed environment and thus has the ability to detect anomalous conditions and trigger appropriate actions.
\end{itemize}

We could register to DCAE to collect the information we needed to develop a security solution as a VNF to enhance the ability of the ONAP platform. We proposed an end-to-end trainable Tree-Shaped Deep Neural Network (TSDNN) model and a Quantity Dependent Backpropagation (QDBP) algorithm at  the IEEE PIMRC in 2017  \cite{b12} to analyze malicious flows and expected it to become one of  the potential security solutions.

\section{5G Potential Security Threats}
We can clearly identify where the potential threats to 5G lie based on the knowledge mentioned above. The following potential threats to the 5G network are as follows.

\subsection{Potential Security Threats in Network Slicing}
3GPP TR 33.811\cite{b13} is a new topic that was created in November 2017 for security issues. Four key security issues with NS and attack tactics have been identified.
\subsubsection{Unauthorized Access to Management Exposure Interface}
Due to the flexibility of the network slice, we can see from Fig.\ref{fig:NS} that there must be an interface between the NG-RAN and vEPC or vEPC and the application service in the cloud. If the interface is subject to unauthorized accessed, the network slice isolation may be invalid and the following attacks may occur.
\begin{itemize}
	\item Services may be changed in unauthorized ways. 
	\item A fake NSI may be created to track customers.
	\item NS services may be deleted or modified to achieve a DoS attack.
	\item The routing mechanism may be changed to achieve a man-in-the-middle attack.
\end{itemize}

\subsubsection{Protecting the Results of NSI Supervision/Reporting}
As with VMware or KVM, a host is needed to integrate the various virtualized parts, and NSI supervision/reporting is also necessary. This secret information must be well protected.
\begin{itemize}
	\item Information might be disclosed to the attacker when the information is not encrypted.
	\item The NS service may be modified by tampering with NSI supervision/reporting.
\end{itemize}

\subsubsection{Protecting Network Slice Subnet Template(NSST)}
Many templates are pre-set in the design of the network slicing to enable rapid response to user demand. The template must be protected to prevent it from being probed.
\begin{itemize}
	\item Attack NSI by collecting information when the information is not encrypted.
	\item Tamper with the NSST to fail the NS.
\end{itemize}

\subsubsection{Insecure Procedure for Capability Negotiation}
Take the OpenSSL downgrade attack as an example. If a man-in-the-middle attack has been successfully performed, the hacker can deliberately inform that it only supports the encryption protocol of the older version that has been cracked. The communication packets are then encrypted by a weaker protocol so the hacker can easily decrypt the encrypted packet and steal the information.
\begin{itemize}
	\item Downgrade the NS capabilities by man-in-the-middle attack.
\end{itemize}

\subsection{Security Threats to NG-RAN}
\subsubsection{DoS Attack}
Malicious base stations interfere with other base stations by sending out useless network traffic signals. This kind of attack can only block the service but cannot steal the data.

\subsubsection{Primary User Simulation Attack}
There are primary users and secondary users in the NG-RAN, and the network environment is more complicated than the traditional core network. For example, a malicious attacker can occupy a specific idle spectrum band by simulating the characteristics of the primary user. When the secondary user wants to implement the spectrum resource, the malicious attacker can respond to the request that there is no idle resource to reject the secondary user.

\subsubsection{Virtual BBU pool security}
Just as the SDN controller controls many SDN switches, the virtual BBU pool manages multiple base stations simultaneously. This also makes it a target for hackers. The virtual BBU pool has a VNF platform. If the hacker enters through the platform interface, several base stations can be affected at the same time. Compared with 4G base stations, which are closed and independent, the security of virtualized BBUs pool is a new topic.

\subsection{Security Threats to the vEPC}
\subsubsection{Security of the Network Exposure function (NEF)}
Because dynamic and flexible deployment of 5G is necessary, there must be an interface for exchanging messages.
The NEF is a new vEPC function that provides an exposure interface to exchange information between internal and external networks. Ensuring the security of the NEF is a new topic. If an exchanged message is spoofed or tampered with, it will cause great harm to the whole vEPC.
\subsubsection{N4 interface security}
The N4 interface connects the UPF and AMF. This interface is closed in a 4G EPC, but not in 5G vEPC, where an encryption mechanism is needed to protect it. Security between the UPF and AMF is an open issue that still needs to be solved.

\subsection{Security Threats to SDN}
\cite{b14} proposes four security threats to SDN.
\subsubsection{Centralized Control} Centralized control presents a  potential target to attackers, who may try to compromise the SDN controller or to control the entire network by tricking.
\subsubsection{Programmability}Because SDN offers clients a clearly programmatic access, so service providers need to pay more attention to system integrity, third-party data, and open interfaces.
\subsubsection{Challenge of Integrating Legacy Protocols}
SDN needs to integrate legacy protocols, but it is difficult to retrofit security ability into existing technologies such as domain name server, border
gateway protocol, and network address translation. It is important to avoid repeating the weaknesses of the existing technologies to in the SDN framework.
\subsubsection{Cross Domain Connection}
One means of SDN implementation is to connect with different domains’ infrastructures. A trust model or authorization mechanism is needed to prevent insecure connections.

\setcounter{subsubsection}{0}
	\cite{b17} also identifies three attacks from the data plane in SDN.
	\subsubsection{Flooding Attack} In the process of network transmission, the sender needs to use the ARP protocol to obtain the MAC address of the receiver so that packets can be correctly forwarded to the receiver via L2 switches. The ARP protocol asks for the MAC address based on the destination IP address, and the sender puts the MAC address into its cache for future use. An attacker can send a flood of fake ARP control messages so that the sender caches incorrect information. Thus, the packet will use the wrong MAC address and be forwarded to the wrong host.
	
	\subsubsection{Host Tracking Service Attack} In a software defined network, the central controller relies on a host tracking service to confirm the host's position in the topology. Applications use this location to calculate routing paths. After a packet is received, the switches report the host's location to the central controller. An attacker keeps sending packets with the IP address of a victim as the source IP address. When the switch connecting to the attacker receives the packets, it notifies the central controller that the victim is connecting to the switch, which is wrong location information and results in wrong path calculation. Thus, all traffic for the victim will be forwarded to the attacker by mistake.
	
	\subsubsection{Link Layer Discovery  Protocol (LLDP) Attack} In a software-defined network, the central controller uses the topology discovery service to obtain the connections among switches and knows the entire network topology. The topology discovery service mainly requires the central controller to broadcast LLDP packets to the switches, and the adjacent switches send LLDP packets back to the central controller. Thus, the central controller knows which switches are adjacent and can calculate the entire topology. Because the LLDP packets are broadcast to the network, an attacker can receive the LLDP packets, then forge them and send them back to the controller. The controller will mistakenly assume that the attacker is the newly added shortest path and further update the path. All packets destined for a victim will pass through the attacker, and thus a man-in-the-middle attack is formed.

More SDN security threats are identified in \cite{b15} \cite{b16}.

\section{Conclusion}
As we approach the 5G era, we need to prepare for new cybersecurity issues. In this paper, we introduce the 5G E2E architecture framework, NG-RAN, vEPC, SDN, NFV, and NFV MANO. We then analyze four MANO implementations and identify opportunities on the ONAP platform where security solutions need to be developed. In this paper, we do not concentrate on the known security threats, but report potential threats to the architecture in the hope that the problems can be solved in the development stage instead of remedying them after attacks take place.

\section*{Acknowledgment}

This work was supported by Minister of Science and Technology, Taiwan under Grant MOST 107-2218-E-002-048 and MOST 108-2634-F-002-009.


\begin{thebibliography}{00}
\bibitem{b1} W. Kiess, M. R. Sama, J. Varga, J. Prade, H. J. Morper and K. Hoffmann, "5G via evolved packet core slices: Costs and technology of early deployments," 2017 IEEE 28th Annual International Symposium on Personal, Indoor, and Mobile Radio Communications (PIMRC), Montreal, QC, 2017, pp. 1-7.
\bibitem{b2} Huawei, ``5G Unlocks A World of Opportunities: Top Ten 5G Use Cases ", Jan. 2018.
\bibitem{b3} 3GPP, ``Study on management and orchestration of network slicing for next generation network", release-15, v. 15.1.0. In TR 28.801, Jan 2018.
\bibitem{b4} 3GPP, ``System architecture for the 5g system; stage 2", release-15, v. 15.1.0. In TS 23.501, Mar 2018.
\bibitem{b5} 3GPP, ``Procedures for the 5g system; stage 2",release-15, v. 15.1.0. In TS 23.502, Mar 2018.
\bibitem{b6} F. Tian, P. Zhang and Z. Yan, "A Survey on C-RAN Security," in IEEE Access, vol. 5, pp. 13372-13386, 2017.
\bibitem{b7} Y. Lin, L. Shao, Z. Zhu, Q. Wang and R. K. Sabhikhi, "Wireless network cloud: Architecture and system requirements," in IBM Journal of Research and Development, vol. 54, no. 1, pp. 4:1-4:12, January-February 2010.
\bibitem{b8} N. McKeown, T. Anderson, H. Balakrishnan, G. Parulkar, L. Peterson, J. Rexford, S. Shenker, and J. Turner. 2008. ``OpenFlow: enabling innovation in campus networks". SIGCOMM Comput. Commun. Rev. 38, 2 (March 2008), 69-74. 
\bibitem{b9} Network Functions Virtualisation (NFV); Architectural Framework, ETSI Standard GS NFV 002, 2014.
\bibitem{b17} ETSI, ``An ETSI OSM Community White Paper", OSM RELEASE THREE, A TECHNICAL OVERVIEW, October 2017
\bibitem{b18} SONATA, ``SONATA NFV: Agile service development and orchestration in 5G
\bibitem{b19}F. Z. Yousaf, M. Bredel, S. Schaller and F. Schneider, ``NFV and SDN—Key Technology Enablers for 5G Networks", in IEEE Journal on Selected Areas in Communications, vol. 35, no. 11, pp. 2468-2478, Nov. 2017.
\bibitem{b10} OPNFV, “Open platform for NFV (OPNFV),” Open Source Project, 2017, accessed: Nov. 2, 2017. [Online]. Available: https://www.opnfv.org/
\bibitem{b11} ONAP, “ONAP Documentation” , Beijing Release, Jun 2018 ,
[Online]. Available: http://onap.readthedocs.io/en/beijing/
\bibitem{b12} Y. Chen, Y. Li, A. Tseng and T. Lin, "Deep learning for malicious flow detection," 2017 IEEE 28th Annual International Symposium on Personal, Indoor, and Mobile Radio Communications (PIMRC), Montreal, QC, 2017, pp. 1-7.
\bibitem{b13} 3GPP, ``study on security aspects of 5g network slicing management ",release-15, v. 1.0.0. In TR 33.811, Jun 2018.
\bibitem{b14} Open Networking Foundation, ``Principles and practices for securing
software-defined networks". Jan 2015.
\bibitem{b17} Sungmin Hong, Lei Xu, Haopei Wang, Guofei Gu, ``Poisoning Network Visibility in Software-Defined Networks: New Attacks and Countermeasures",NDSS Symposium 2015, San Diego, 2015
\bibitem{b15} I. Ahmad, T. Kumar, M. Liyanage, J. Okwuibe, M. Ylianttila and A. Gurtov, ``5G security: Analysis of threats and solutions", 2017 IEEE Conference on Standards for Communications and Networking (CSCN), Helsinki, 2017, pp. 193-199.
\bibitem{b16} Q. Yan, F. R. Yu, Q. Gong and J. Li, ``Software-Defined Networking (SDN) and Distributed Denial of Service (DDoS) Attacks in Cloud Computing Environments: A Survey, Some Research Issues, and Challenges", in IEEE Communications Surveys and Tutorials, vol. 18, no. 1, pp. 602-622, Firstquarter 2016.

\end{thebibliography}
\end{document}